\documentclass[12pt,onecolumn]{revtex4}
\usepackage{amssymb,epsf}
\begin{document}

\title{Thermodynamics of Rotating Charged Black Strings\\
and (A)dS/CFT Correspondence}
\author{M. H. Dehghani}\email{dehghani@physics.susc.ac.ir}
\address{Physics Department and Biruni  Observatory,\\
         Shiraz University, Shiraz 71454, Iran\\ and\\
         Institute for Studies in Theoretical Physics and Mathematics (IPM)\\P.O. Box 19395-5531, Tehran, Iran}
\begin{abstract}
We use the anti-de Sitter (AdS) conformal field theory
correspondence to calculate the conserved charges and the
Euclidean actions of the charged rotating black string in four
dimensions both in the canonical and the grand-canonical ensemble.
The four-dimensional solution of the Einstein-Maxwell equations
with a positive cosmological constant is introduced, and its
conserved quantities and actions for fixed charge and fixed
electric potential are calculated. We also study the
thermodynamics of the asymptotically AdS black strings and perform
a stability analysis both in the canonical and the grand-canonical
ensembles. We find that the asymptotically AdS black string in the
canonical ensemble is locally stable, while in the grand-canonical
ensemble it is stable only for a part of the phase space.
\end{abstract}

\maketitle


\section{Introduction\label{Intro}}

The thermodynamics of asymptotically anti-de Sitter (AdS) black
holes continues to attract a great deal of attention. This is due
to the fact that there is a correspondence between super gravity
(the low energy limit of string theory) in $(n+1)$-dimensional
asymptotically AdS spacetimes and conformal field theory (CFT)
living on an $n$-dimensional boundary known as the AdS/CFT
correspondence \cite{Mal1}. Indeed, one can gain some insight into
the thermodynamic properties and phase structures of strong 't
Hooft coupling conformal field theories by studying the
thermodynamics of asymptotically AdS black holes. A standard
example of the AdS/CFT correspondence is the interpretation of the
Hawking-Page phase transition from a low-temperature confining to
a high-temperature deconfining phase in the dual field theory
\cite{Wit1}.

The AdS/CFT correspondence is now a fundamental concept which
furnish a means for calculating the action and thermodynamical
quantities intrinsically without reliance on any reference
spacetime \cite {BK1,Hen,Emp,Mann}. This conjecture has been
recently extended to the case of asymptotically de Sitter
spacetimes \cite{Stro1,Deh1}. Although the (A)dS/CFT
correspondence applies for the case of spacially infinite
boundary, it was also employed for the computation of the
conserved and thermodynamic quantities in the case of a finite
boundary \cite{Deh2}. This conjecture has also been used for the
case of black objects with nonspherical horizons \cite{Deh3}.

It is well known that the Einstein equation with positive or
negative cosmological constant has black hole solutions with
horizons being positive, zero, or negative constant curvature
hypersurfaces \cite{Man2}. The rotating solutions of the Einstein
equation with a negative cosmological constant with cylindrical
and toroidal horizons have been studied in Ref. \cite{Lem1}. The
extension to include the Maxwell field has been done and the
static and rotating electrically charged black string have been
considered in Ref. \cite{Lem2}. Recently, asymptotically anti-de
Sitter spacetimes generated by static and spinning magnetic string
sources in general relativity have been also considered
\cite{Lem3}. The thermodynamics of asymptotically anti-de Sitter
spacetimes with nonspherical horizon has been studied by many
authors \cite{Pec,Emp}. In this paper, we want to apply the
AdS/CFT correspondence to the asymptotically anti-de Sitter
charged rotating black string in four dimensions with cylindrical
or toroidal event horizons, and study their thermodynamics in both
the canonical and the grand-canonical ensemble. The stability
conditions are investigated and a complete phase diagram is
obtained. We also introduce the asymptotic de Sitter charged
rotating black strings with cylindrical and toroidal horizon and
calculate their conserved quantities and Euclidean actions in both
the canonical and the grand-canonical ensemble.

The outline of our paper is as follows. We review the basic
formalism in Sec. \ref{Action}. In Sec. \ref{BString} we consider
the four-dimensional charged rotating black strings which are
asymptotically anti-de Sitter and introduce the asymptotically de
Sitter charged rotating strings. We also compute the conserved
charges and the Euclidean actions for asymptotically AdS and dS
strings. In Sec. \ref{Therm}, we study the thermodynamics of the
string in both the canonical and the grand-canonical ensemble, and
the thermal stability of the black strings is investigated. We
finish our paper with some concluding remarks.

\section{The Action and Conserved Quantities\label{Action}}

The gravitational action of four-dimensional asymptotically
(anti)-de Sitter spacetimes $\mathcal{M}$, with boundary $\delta
\mathcal{M}$ in the presence of an electromagnetic field is
\begin{equation}
I_G=-\frac 1{16\pi }\int_{\mathcal{M}}d^4x\sqrt{-g}\left( \mathcal{R}\text{ }%
-2\Lambda -F_{\mu \nu }F^{\mu \nu }\right) +\frac 1{8\pi }\int_{\partial \mathcal{M}}d^3x\sqrt{%
-\gamma }\Theta (\gamma ),  \label{Actg}
\end{equation}
where $F_{\mu \nu }=\partial _\mu A_\nu -\partial _\nu A_\mu $ is
the electromagnetic tensor field and $A_\mu $ is the vector
potential. The first term is the Einstein-Hilbert volume term with
negative (AdS) or positive (dS) cosmological constant $\Lambda
=\mp 3/l^2$, and the second term is the Gibbons-Hawking boundary
term which is chosen such that the variational principle is
well-defined. The
manifold $\mathcal{M}$ has metric $g_{\mu \nu }$ and covariant derivative $%
\nabla _\mu $. $\Theta $ is the trace of the extrinsic curvature $\Theta
^{\mu \nu }$ of any boundary(ies) $\partial \mathcal{M}$ of the manifold $%
\mathcal{M}$, with induced metric(s) $\gamma _{ij}$. In general
the first and second terms of Eq. (\ref{Actg}) are both divergent
when evaluated on the solutions, as is the Hamiltonian, and other
associated conserved quantities. Rather than eliminating these
divergences by incorporating a reference term in the spacetime
\cite{BY,BCM}, a new term, $I_{ct}$, is added to the action which
is a functional only of the boundary curvature invariants. The
counterterms for asymptotically AdS and dS spacetimes in four
dimensions are \cite{BK1,Deh1}:
\begin{equation}
I_{ct}=-\frac 1{4\pi l}\int_{\partial \mathcal{M}_\infty }d^3x\sqrt{-\gamma }%
\left( 1\pm \frac{l^2}4R\right) ,  \label{Actct}
\end{equation}
where $R$ is the Ricci scalar of the boundary metric $\gamma
_{ab}$. The $+$ and $-$ sign correspond to asymptotically AdS and
dS spacetimes, respectively. These counterterms have been used by
many authors for a wide variety of the spacetimes, including
Schwarzschild-(A)dS, topological Schwarzschild-(A)dS, Kerr-(A)dS,
Taub-NUT (Newmann-Unti-Tamborino)-AdS, Taub-bolt-AdS, and
Taub-bolt-Kerr-AdS \cite{Emp,Mann,Deh1}. The total action can be
written as a linear combination of the gravity term (\ref{Actg})
and the counterterms (\ref{Actct}) as
\begin{equation}
I=I_G+I_{ct}.  \label{totact}
\end{equation}
In order to obtain the Einstein-Maxwell field equations by the variation of
the volume integral with respect to the fields, one should impose the
boundary condition $\delta A_\mu =0$ on $\delta \mathcal{M}$. Thus the
action (\ref{totact}) is appropriate to study the grand-canonical ensemble
with fixed electric potential \cite{Cal}.

To study the canonical ensemble with fixed electric charge, one should
impose the boundary condition $\delta (n^aF_{ab})=0$, and therefore the
total action is \cite{Haw1}
\begin{equation}
\tilde{I}=I-\frac 1{4\pi }\int_{\partial \mathcal{M}_\infty }d^3x\sqrt{%
-\gamma }n_aF^{ab}A_b.  \label{Actb}
\end{equation}
Using the Brown and York definition \cite{BY} one can construct a
divergence-free stress-energy tensor from the total action
(\ref{totact}) as
\begin{equation}
T^{ab}=\frac 1{8\pi }\left\{ (\Theta ^{ab}-\Theta \gamma ^{ab})-\frac
2l\left[ \gamma ^{ab}\pm \frac{l^2}2\left( R^{ab}-\frac 12R\gamma
^{ab}\right) \right] \right\} .  \label{Stres}
\end{equation}
Again the $+$ and $-$ sign correspond to asymptotically AdS and dS
spacetimes, respectively.

To compute the conserved charges of the spacetime, one should
choose a
spacelike surface $\mathcal{B}$ in $\partial \mathcal{M}$ with metric $%
\sigma _{ij}$, and write the boundary metric in ADM form:
\[
\gamma _{ab}dx^adx^a=-N^2dt^2+\sigma _{ij}\left( d\phi ^i+V^idt\right)
\left( d\phi ^j+V^jdt\right) ,
\]
where the coordinates $\phi ^i$ are the angular variables
parametrizing the hypersurface of constant $r$ around the origin.
When there is a Killing vector field $\mathcal{\xi }$ on the
boundary, then the conserved quantities associated with the stress
tensors of Eq. (\ref{Stres}) can be written as
\begin{equation}
\mathcal{Q}(\mathcal{\xi )}=\int_{\mathcal{B}_\infty }d^{n-1}\phi \sqrt{%
\sigma }T_{ab}n^a\mathcal{\xi }^b,  \label{charge}
\end{equation}
where $\sigma $ is the determinant of the metric $\sigma _{ij}$, $\mathcal{%
\xi }$ and $n^a$ are the Killing vector field and the unit normal
vector on the boundary $\mathcal{B}$ . For boundaries with
timelike ($\xi =\partial /\partial t$) and rotational Killing
vector fields ($\varsigma =\partial /\partial \phi $), we obtain
\begin{eqnarray}
M &=&\int_{\mathcal{B}_\infty }d^2\phi \sqrt{\sigma }T_{ab}n^a\xi ^b,
\label{Mastot} \\
J &=&\int_{\mathcal{B}_\infty }d^{n-1}\phi \sqrt{\sigma }T_{ab}n^a\zeta ^b,
\label{Angtot}
\end{eqnarray}
provided the surface $\mathcal{B}$ contains the orbits of $\zeta$.
These quantities are, respectively, the conserved mass and angular
momentum of the system enclosed by the boundary. Note that they
will both be dependent on the location of the boundary
$\mathcal{B}$ in the spacetime,
although each is independent of the particular choice of foliation $\mathcal{%
B}$ within the surface $\partial \mathcal{M}$.

In the context of the (A)dS/CFT correspondence, the limit in which
the boundary $\mathcal{B}$ becomes infinite ($\mathcal{B}_\infty
$) is taken, and the counterterm prescription ensures that the
action and conserved charges are finite. No embedding of the
surface $\mathcal{B}$ into a reference spacetime is required and
the quantities which are computed are intrinsic to the spacetimes.

\section{Rotating Charged Black Strings in Four dimensions\label{BString}}

In this section we consider the solutions of the Einstein-Maxwell
equations with negative and positive cosmological constants which
possess cylindrical symmetry. We see that for a suitable choice of
the parameters, these solutions describe stationary charged
rotating black strings with cylindrical or toroidal horizons.

\subsection{Asymptotically anti-de Sitter case}

The asymptotically AdS solution of the Einstein-Maxwell equations
with cylinderical symmetry can be written as \cite{Lem2}
\begin{eqnarray}
ds^2 &=&-\Xi ^2\left( f(r)-\frac{a^2r^2}{\Xi ^2l^4}\right)
dt^2+\frac 1{f(r)}dr^2-2 \frac{a\Xi l}r\left( b-\frac lr\lambda
^2\right) dt d\phi
\nonumber \\
&&+\left[ \Xi ^2r^2-a^2f(r)\right] d\phi ^2+\frac{r^2}{l^2}dz^2,
\label{met1a}
\end{eqnarray}
\begin{equation}
A_\mu =-\Xi \frac{l \lambda }r(\delta _\mu ^0-\frac a\Xi \delta
_\mu ^2), \label{met1b}
\end{equation}
where
\begin{eqnarray}
f(r) &=&\frac{r^2}{l^2}-\frac{bl}r+\frac{\lambda ^2l^2}{r^2},  \nonumber \\
\Xi ^2 &=&1+\frac{a^2}{l^2}.  \label{met1c}
\end{eqnarray}
$a$, $b$, and $\lambda $ are the constant parameters of the
metric. It is worthwhile to mention that for the case of $-\infty
<z<\infty $, Eqs. (\ref {met1a})-(\ref{met1c}) describe a
stationary black string with cylindrical horizon, and if one
compactifies the $z$ coordinate ($0\leq z<2\pi l$) one has a
closed black string (or black hole) with toroidal horizon. As we
will show in the next section the angular momentum is proportional
to the parameter $a$, and therefore $a$ is the rotational
parameter of the spacetime. It is easy to show that in the
nonrotating case ($a=0$), $\lambda /2$ and $b/4$ are the linear
charge and mass densities of the $z$-line as we will see in the
next section.

The metric of Eqs. (\ref{met1a})-(\ref{met1c}) has two inner and
outer horizons located at $r_{-}$ and $r_{+}$, provided the
parameter $b$ is greater than $b_{crit}$ given as
\begin{equation}
b_{crit}=4\times 3^{-3/4}\lambda ^{3/2}. \label{bcrit}
\end{equation}
In the case that $b=b_{crit}$, we will have an extreme black
string. The horizon area per unit length of the string for the
case of a cylindrical horizon is $2\pi \Xi r_{+}^2/l$. Since the
area law of entropy is universal, and applies to all kinds of
black holes and black strings \cite {Beck,Haw1,HH,Mann}, the
entropy per unit length is
\begin{equation}
\mathcal{S}=\frac{\pi \Xi r_{+}^2}{2l}.  \label{Ent}
\end{equation}
For the case of a toroidal horizon, the horizon area of the string is $2\pi l%
\mathcal{S}$.

Analytical continuation of the Lorentzian metric by $t\rightarrow i\tau $
and $a\rightarrow ia$ yields the Euclidean section, whose regularity at $%
r=r_{+}$ requires that we should identify $\tau \sim \tau +\beta _{+}$ and $%
\phi \sim \phi +i\beta _{+}\Omega _{+}$, where $\beta _{+}$ and
$\Omega _{+}$ are the inverse Hawking temperature and the angular
velocity of the outer event horizon. It is a matter of calculation
to show that
\begin{eqnarray}
\beta _{+} &=&\frac{4\pi \Xi }{f^{\prime }(r_{+})}=\frac{4\pi \Xi l^2r_{+}^3%
}{3r_{+}^4-\lambda ^2l^4},  \label{bet1} \\
\Omega _{+} &=&\frac a{\Xi l^2}.  \label{Om1}
\end{eqnarray}

Using Eqs. (\ref{Actg})-(\ref{Actb}) the Euclidean action per unit
length of the black string with a cylindrical horizon in the
grand-canonical and the canonical ensemble can be calculated as
\begin{eqnarray}
\mathcal{I} &=&-\frac{\beta _{+}}8b,  \label{Act1ads} \\
\widetilde{\mathcal{I}} &=&\frac{\beta _{+}}8
\left(3b-4\frac{r_{+}^3}{l^3}\right), \label{Act2ads}
\end{eqnarray}
valid for a fixed potential and a fixed charge, respectively. For
the case of
toroidal black strings the total action is $2\pi l\mathcal{I}$ and $%
2\pi l\widetilde{\mathcal{I}}$, respectively.

The conserved mass and angular momentum per unit length of the
string with a cylinderical horizon calculated on the boundary
$\mathcal{B}$ at infinity can be calculated through the use of
Eqs. (\ref{Mastot}) and (\ref{Angtot}),
\begin{equation}
 \mathcal{M}=\frac 18(3\Xi ^2-1)b,\hspace{.5cm }\mathcal{J}=\frac
38\Xi ba.  \label{Jads}
\end{equation}
For $a=0$ ($\Xi=1$), the angular momentum and mass per unit length
are $0$ and $b/4$, respectively. Thus $a$ is the rotational
parameter and $b/4$ is associated to the mass density per unit
length. The total mass and
angular momentum of the string with toroidal horizon are $2\pi l\mathcal{M}%
$ and $2\pi l\mathcal{J}$, respectively.

The charge per unit length, $\mathcal{Q}$, can be found by
calculating the flux of the electromagnetic field at infinity,
yielding
\begin{equation}
\mathcal{Q}=\frac{\Xi \lambda }2.  \label{chden}
\end{equation}
The electric potential $\Phi $, measured at infinity with respect
to the horizon, is defined by \cite{Cal}
\begin{equation}
\Phi =A_\mu \chi ^\mu \left| _{r\rightarrow \infty }-A_\mu \chi
^\mu \right| _{r=r_{+}},  \label{Pot}
\end{equation}
where $\chi =\partial _t+\Omega _{+}\partial _\phi $ is the null
generator of the horizon. One finds
\begin{equation}
\Phi = \frac{\lambda l}{\Xi r_{+}}.  \label{Pot1}
\end{equation}

\subsection{Asymptotically de Sitter case\label{dS}}

The solution of the Einstein-Maxwell equations with a positive
cosmological constant which has cylindrical symmetry can be
written as
\begin{eqnarray}
ds^2 &=&-\Gamma ^2\left( h(r)-\frac{a^2r^2}{\Gamma ^2l^4}\right)
dt^2+\frac 1{h(r)}dr^2-2 \frac{a\Gamma l}r\left( b-\frac lr\lambda
^2\right) dtd\phi
\nonumber \\
&&+\left[ \Gamma ^2r^2-a^2h(r)\right] d\phi ^2+\frac{r^2}{l^2}dz^2,
\label{met2a}
\end{eqnarray}
\begin{equation}
A_\mu =-i \Gamma \frac{l \lambda }r(\delta _\mu ^0+\frac a\Gamma
\delta _\mu ^2), \label{met2b}
\end{equation}
where
\begin{eqnarray}
h(r) &=&-\left( \frac{r^2}{l^2}-\frac{bl}r+\frac{\lambda ^2l^2}{r^2}\right) ,
\nonumber  \label{met1b} \\
\Gamma ^2 &=&1-\frac{a^2}{l^2}.  \label{met2c}
\end{eqnarray}
Again $a$, $b$, and $\lambda $ are the constant parameters of the
metric.
For a suitable choice of the parameters $a$, $b$, and $\lambda $, the solution (%
\ref{met2a})-(\ref{met2c}) describes a stationary black string
with cylindrical or toroidal horizons. In the case of $-\infty
<z<\infty $, one has a black string with a cylindrical horizon,
and if one compactifies the $z$ coordinate ($0\leq z<2\pi l$) one
has a closed black string (or black hole) with a toroidal horizon.
Again, as in the case of asymptotically anti-de Sitter spacetimes,
$a$, $b/4$, and $\lambda /2$ can be interpreted as the rotational
parameter, the mass, and charge densities of the $z$-line.

The metric of Eqs. (\ref{met2a})-(\ref{met2c}) has two outer and
cosmological horizons located at $r_{+}$ and $r_c$ provided the
parameter $b$ is greater than $b_{crit}$ given by Eq.
(\ref{bcrit}). In the case in which $b=b_{crit}$, we will have an
extreme black string. The area law of the entropy also applies to
the cosmological event horizon of the asymptotic de Sitter black
holes \cite{GH1}, and therefore the entropy of the string for the
case of cylindrical and toroidal horizons is $\mathcal{S}_{ds}=\pi
\Gamma r_{+}^2/(2l)$ and $2\pi l \mathcal{S}_{ds}$, respectively.

Again the inverse Hawking temperature and the angular velocity of the
cosmological event horizon can be calculated as
\begin{eqnarray}
\beta _c &=&\frac{4\pi \Gamma }{h^{\prime }(r_c)}=-\frac{4\pi \Gamma l^2r_c^3}{%
3r_c^4-\lambda ^2l^4},  \label{bet2} \\
\Omega _c &=&\frac a{\Gamma l^2}.  \label{Om2}
\end{eqnarray}

Using Eqs. (\ref{Actg})-(\ref{Actb}) for the de Sitter case, the
Euclidean action per unit length of the black string with a
cylindrical horizon in the grand-canonical and the canonical
ensemble can be calculated as
\begin{eqnarray}
\mathcal{I}_{ds} &=&\frac{\beta _c}8b,  \label{Act1ds} \\
\widetilde{\mathcal{I}}_{ds} &=&-\frac{\beta _c}8(3b-4\frac{r_c^3}{l^3}).
\label{Act2d}
\end{eqnarray}
For the case of a toroidal black string the total action in the
grand-canonical
and the canonical ensemble is $2\pi l\mathcal{I}_{ds}$ and $2\pi l%
\widetilde{\mathcal{I}}_{ds}$.

The conserved mass and angular momentum per unit length of the
string with a cylindrical horizon calculated on the boundary
$\mathcal{B}$ at infinity can be calculated through the use of
Eqs. (\ref{Mastot}) and (\ref{Angtot}),
\begin{equation}
\mathcal{M}_{ds}=-\frac 18(3\Gamma ^2-1)b,\hspace{.5cm
}\mathcal{J}_{ds}=\frac 38\Gamma ba.  \label{Jds}
\end{equation}
For $a=0$, the angular momentum and mass per unit length are $0$
and $b/4$, respectively. Thus the parameters $a$ and $b/4$ are
associated to the angular momentum and mass density per unit
length. The total mass and
angular momentum of the string with toroidal horizon are $2\pi l\mathcal{M}%
_{ds} $ and $2\pi l\mathcal{J}_{ds}$, respectively.

Here the charge per unit length is $\mathcal{Q}_{ds}=\Gamma
\lambda /2$, and the electric potential $\Phi $ can be calculated
as
\begin{equation}
\Phi_{ds} =-\frac{\lambda l}{\Gamma r_c}.  \label{Pot2}
\end{equation}

\section{Thermodynamics of Black Strings}\label{Therm}
\subsection{Energy as a function of entropy, angular momentum, and charge}
We first obtain the mass per unit length as a function of $S,$
$J$, and $Q$. Using the expression (\ref{Jads}) for the mass and
angular momentum per unit length, Eq. (\ref{Ent}) for the entropy,
and the fact that $f(r_{+})=0$, one obtains by simple algebraic
manipulation
\begin{eqnarray}
\mathcal{M} &=&(8\pi ^3l^3\mathcal{S})^{-1/2}\left\{ \Upsilon -27\frac{(%
\mathcal{S}^2+l^2\pi ^2\mathcal{Q}^2)^4}\Upsilon \right\} ,  \nonumber
\label{Mas} \\
\Upsilon &=&4\pi ^3l\mathcal{SJ}^2+\sqrt{16\pi ^6l^2\mathcal{S}^2\mathcal{J}%
^4+81(\mathcal{S}^2+l^2\pi ^2\mathcal{Q}^2)^4.}  \label{Mass}
\end{eqnarray}
One may then regard the parameters $\mathcal{S}$, $\mathcal{J}$, and $%
\mathcal{Q}$ as a complete set of energetic extensive parameters
for the mass per unit length
$\mathcal{M}=\mathcal{M}(\mathcal{S},\mathcal{J},\mathcal{Q})$ and
define the quantities conjugate to $\mathcal{S},$ $\mathcal{J}$,
and $\mathcal{Q}$. These quantities are the temperature, the
angular velocity, and the electric potential
\begin{equation}
T=\left( \frac{\partial \mathcal{M}}{\partial \mathcal{S}}\right) _{\mathcal{JQ}%
},\hspace{.5cm }\Omega =\left( \frac{\partial \mathcal{M}}{\partial \mathcal{J}}%
\right) _{\mathcal{SQ}},\hspace{.5cm }\Phi =\left( \frac{\partial \mathcal{M}}{%
\partial \mathcal{Q}}\right) _{\mathcal{JS}}.  \label{TMP}
\end{equation}
It is a matter of straightforward calculation to show that the
quantities calculated by Eq. (\ref{TMP}) for the temperature, the
angular velocity, and the electric potential coincide with Eqs.
(\ref{bet1}), (\ref{Om1}), and (\ref {Pot1}) found in Sec.
\ref{BString}. Thus, the thermodynamical quantities calculated in
Sec. \ref{BString} satisfy the first law of thermodynamics,
\begin{equation}
d\mathcal{M}=Td\mathcal{S}+\Omega d\mathcal{J}+\Phi d\mathcal{Q}.
\label{1law}
\end{equation}

\subsection{Thermodynamic potentials}
We now consider the thermodynamic potentials in the
grand-canonical and the canonical ensemble. For the
grand-canonical ensemble using the definition of the Gibbs
potential $G(T,\Omega ,\Phi )=\mathcal{I}/\beta _{+}$, and the
expression (\ref{Act2ads}) for the action, Eq. (\ref{bet1}) for
the inverse Hawking temperature, Eq. (\ref{Om1}) for the angular
velocity, and Eq. (\ref{Pot1}) for the electric potential of
asymptotically AdS spacetime, we obtain
\begin{equation}
G(T,\Omega ,\Phi )=\mathcal{M}-\Omega
\mathcal{J}-T\mathcal{S}-\Phi \mathcal{Q}, \label{Gib}
\end{equation}
which means that $G(T,\Omega ,\Phi )$ is indeed the Legendre
transformation of the energy
$\mathcal{M}(\mathcal{S},\mathcal{J},\mathcal{Q})$ with respect to
$\mathcal{S}$, $\mathcal{J}$, and $\mathcal{Q}$. It is a matter of
straightforward calculations to show that the extensive quantities
\[
\mathcal{S}=-\left( \frac{\partial G}{\partial T}\right) _{\Omega \Phi },%
\hspace{.5cm }\mathcal{J}=-\left( \frac{\partial G}{\partial \Omega }\right)
_{T\Phi },\hspace{.5cm }\mathcal{Q}=-\left( \frac{\partial G}{\partial \Phi }%
\right) _{T\Omega },
\]
turn out to coincide precisely with the expressions (\ref{Ent}), (\ref{Jads}%
), and (\ref{chden}).

For the canonical ensemble, the Helmholtz free energy $F(T,\mathcal{J},%
\mathcal{Q})$ is defined as
\begin{equation}
F(T,\mathcal{J},\mathcal{Q})=\frac{\widetilde{\mathcal{I}}}\beta +\Omega
\mathcal{J},  \label{Hel1}
\end{equation}
where $\widetilde{\mathcal{I}}$ is given by Eq. (\ref{Act2ads}).
One can verify that the conjugate quantities
\[
\mathcal{S}=-\left( \frac{\partial F}{\partial T}\right) _{\mathcal{JQ}},%
\hspace{.5cm }\Omega =\left( \frac{\partial F}{\partial \mathcal{J}}\right)
_{T\mathcal{Q}},\hspace{.5cm }\Phi =\left( \frac{\partial F}{\partial
\mathcal{Q}}\right) _{T\mathcal{J}},
\]
agree with expressions (\ref{Ent}), (\ref{Om1}), and (\ref{Pot1})
and one has also
\begin{equation}
F(T,\mathcal{J},\mathcal{Q})=\mathcal{M}-T\mathcal{S}. \label{Hel}
\end{equation}
Thus, $F$ is the Legendre transform of $\mathcal{M}(\mathcal{S},\mathcal{J},%
\mathcal{Q})$ with respect to $\mathcal{S}$.

\subsection{Stability in the canonical and the grand-canonical ensemble}

The local stability analysis in any ensemble can in principle be
carried out by finding the determinant of the Hessian matrix
$[\partial ^2S/\partial X_i\partial X_j]$, where $X_i$'s are the
thermodynamic variables of the system \cite{Cev}. In our case the
entropy $S$ is a function of the mass, the angular momentum, and
the charge per unit length. But, the number of the thermodynamic
variables depends on the ensemble which is used. The more $X_i$ we
regard as variable parameters, the smaller is the region of
stability. In the canonical ensemble, the charge and angular
momentum are fixed parameters, and for this reason the positivity
of the thermal capacity $C_{\mathcal{J},\mathcal{Q}}$ is
sufficient to assure the local stability. The thermal capacity $C_{\mathcal{J%
},\mathcal{Q}}$ at constant charge and angular momentum is
\begin{equation}
C_{\mathcal{J},\mathcal{Q}}=T\frac{\partial \mathcal{S}}{\partial T},
\label{Hcap}
\end{equation}
where $T$ is the inverse of $\beta _{+}$ given by Eq. (\ref{bet1}) and $%
\partial \mathcal{S}/\partial T$ can be calculated as
\begin{equation}
\frac{\partial \mathcal{S}}{\partial T}=\frac{2l\pi ^2\Xi ^2r_{+}^5(\Xi
^2+1)(r_{+}^4+q^2l^4)}{3[4r_{+}^8(\Xi ^2-1)+\Xi
^2(q^2l^4-r_{+}^4)^2+4q^2r_{+}^4l^4]}.  \label{dST}
\end{equation}
As one can see from Eq. (\ref{dST}), $\partial
\mathcal{S}/\partial T$ is positive for all the allowed values of
the metric parameters discussed in Sec. \ref {BString}, and
therefore the asymptotically AdS charged rotating black string in
the canonical ensemble is locally stable.

\begin{figure}
  \epsfxsize=10cm
  \centerline{\epsffile{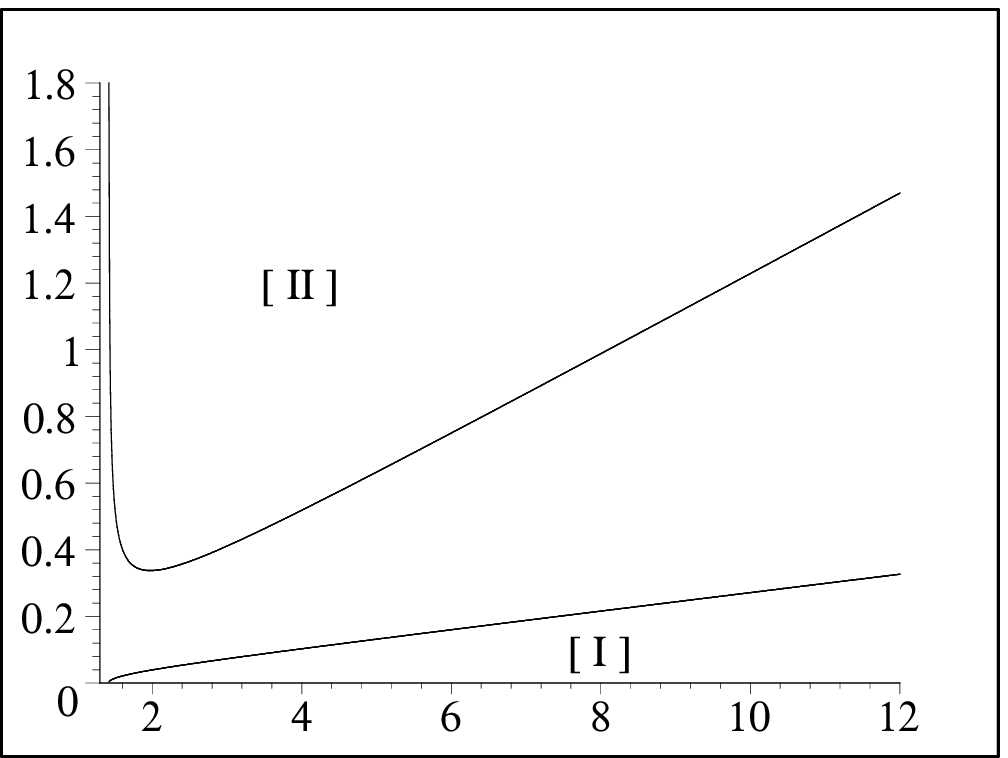}}
  \caption{ $r_+$
versus $\Xi$ for $l=1$ and $\Phi=0.1$.}
  \label{Figure1}
\end{figure}

\begin{figure}
  \epsfxsize=10cm
  \centerline{\epsffile{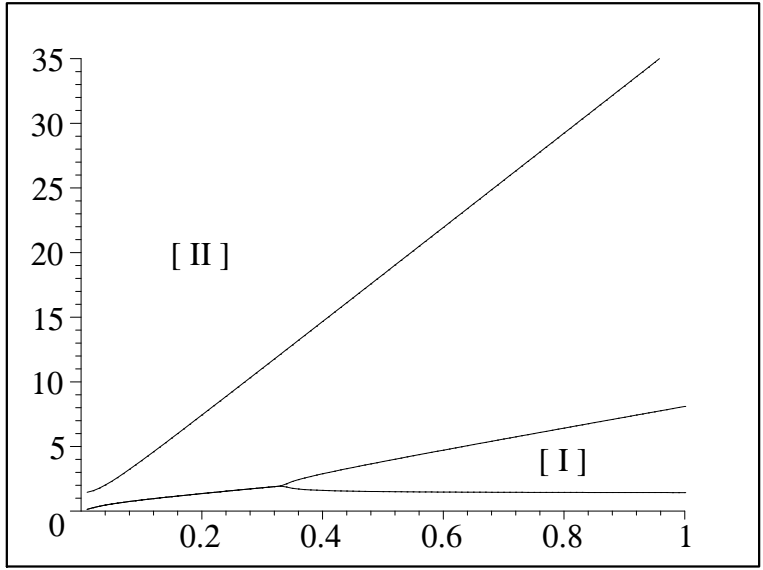}}
  \caption{ $\Xi$
versus $r_+$ for $l=1$ and $\Phi=0.1$.}
  \label{Figure2}
\end{figure}

In the grand-canonical ensemble, the thermodynamic variables are
the mass, the charge and the angular momentum per unit length.
Direct computation of the elements of the Hessian matrix of
$\mathcal{S}(\mathcal{M},\mathcal{J},\mathcal{Q})$ with respect to
$\mathcal{M}$, $\mathcal{J}$, and $\mathcal{Q}$ is a burdensome
task. We find it more efficient to work with the thermodynamic
potential, $G(T,\Omega ,\Phi )$. It is a matter of calculation to
show that the black string is locally stable in the
grand-canonical ensemble if:
\begin{equation}
(\Xi ^2-2)(3r_{+}^4-q^2l^4)^2-8(\Xi ^2+1)q^2l^4r_{+}^4\geq 0.  \label{Gcon}
\end{equation}
As one may note the condition (\ref{Gcon}) is not satisfied for
$\Xi \leq \sqrt{2}$ [or using Eq. (\ref{met1c}) for $a\leq l$].
Thus, the black string is not locally stable for $a\leq l$. For
$\Xi > \sqrt{2}$, we consider the stability condition for a fixed
potential in the ($\Xi ,r_{+}$) plane. The stability condition
(\ref{Gcon}) in terms of the electric potential can be written as
\begin{equation}
9(\Xi ^2-2)r_{+}^4-2l\Phi \Xi ^2(7\Xi ^2-2)r_{+}^2+l^4\Phi ^4\Xi ^4((\Xi
^2-2)\geq 0.  \label{Stab}
\end{equation}
The condition (\ref{Stab}) is satisfied if the radius of the
horizon $r_{+}$ is less than $r_{+1}$ or greater than $r_{+2}$,
where
\begin{eqnarray}
r_{+1} &=& \frac{l\Xi
\Phi}{3 \sqrt{\Xi^2-2}} \{7 \Xi^2-2+ 2\sqrt{10 \Xi^4+2 \Xi^2-8}\}^{1/2}, \label{rh1}\\
r_{+2}&=& \frac{l\Xi \Phi}{3 \sqrt{\Xi^2-2}} \{7 \Xi^2-2-2
\sqrt{10 \Xi^4+2 \Xi^2-8}\}^{1/2}. \label{rh2}
\end{eqnarray}
Figures \ref{Figure1} and \ref{Figure2} show the phase diagrams in
the ($\Xi ,r_{+}$) plane. The black string is locally stable in
the regions [I] and [II] of Figs \ref{Figure1} and \ref{Figure2}.

\section{Closing Remarks}
In this paper, we used the AdS/CFT correspondence to calculate the
conserved quantities and the Euclidean actions for the two cases
of fixed charge and fixed electric potential of asymptotically
anti-de Sitter rotating charged black string. The asymptotically
de Sitter charged rotating black string was introduced and the
conserved quantities and the Euclidean actions were calculated. We
obtained the mass per unit length as a function of the extensive
parameters $\mathcal{S}$, $\mathcal{J}$, and $\mathcal{Q}$,
calculated the temperature, the angular velocity, and the electric
potential, and showed that these quantities satisfy the first law
of thermodynamics. Using the conserved quantities and the
Euclidean actions, the thermodynamic potentials of the system in
the canonical and the grand-canonical ensemble were calculated. We
found that the Helmholtz free energy,
$F(T,\mathcal{J},\mathcal{Q})$, is a Legendre transformation of
the mass per unit length
$\mathcal{M}(\mathcal{S},\mathcal{J},\mathcal{Q})$ with respect to
$\mathcal{S}$, and in the grand-canonical ensemble, the Gibbs
potential is a Legendre transformation of the mass per unit length
with respect to the extensive parameters $\mathcal{S}$,
$\mathcal{J}$ and $\mathcal{Q}$.

The local stability of the asymptotically charged rotating black
string in both the canonical ensemble and the grand-canonical
ensemble was investigated through the use of the Hessian matrix of
the entropy with respect to its thermodynamic variables. We showed
that for the canonical ensemble, where the only thermodynamic
variable was the mass per unit length, the string is locally
stable for the whole phase space. In the grand-canonical ensemble,
where the thermodynamic variables were $\mathcal{S}$,
$\mathcal{J}$, and $\mathcal{Q}$, we showed that the region for
which the string is stable is smaller than for the canonical
ensemble. We found that the black string is unstable for $\Xi \leq
\sqrt{2}$ ($a \leq l$), and for $a > l$ there exist two locally
stable phases. Indeed we studied the thermodynamics of the string
for a fixed electric potential in the ($\Xi ,r_{+}$) plane and
found that the string is locally stable for $r_{+}<r_{+1}$ and
$r>r_{+2}$, where $r_{+1}$ and $r_{+2}$ are given by Eqs.
(\ref{rh1}) and (\ref{rh2}).

The thermodynamics of the asymptotically de Sitter case may be
investigated by the conserved quantities and the Euclidean actions
obtained in Sec. \ref{dS}. But, because of the cosmological
horizon, a more efficient way of its consideration is through the
use of quasilocal thermodynamics, which we give elsewhere.

\end{document}